\documentclass[%
 twocolumn,
 reprint,
 floatfix,
 amsmath,amssymb,
]{revtex4-1}

\usepackage{graphicx}
\usepackage{dcolumn}
\usepackage{bm}

\begin{document}

\preprint{AIP/123-QED}
\title[Structural evolution of protein-biofilms: Simulations and Experiments]{Structural evolution of protein-biofilms: Simulations and Experiments}

\author{Y. Schmitt, H. H\"ahl, C. Gilow, H. Mantz and K. Jacobs}
\affiliation{
Department of Experimental Physics, Saarland University, 66041 Saarbr\"ucken, Germany
}%

\author{O. Leidinger, M. Bellion and L. Santen}
\affiliation{%
Department of Theoretical Physics, Saarland University, 66041 Saarbr\"ucken, Germany
}%

\date{\today}

\begin{abstract}
The control of biofilm formation is a challenging goal that has not been reached yet in many aspects. One is the role of van der Waals forces and another the importance of mutual interactions between the adsorbing and the adsorbed biomolecules ('critical crowding'). Here, a combined exeperimental and theoretical approach is presented that fundamentally probes both aspects. On three model proteins, lysozyme, $\alpha$-amylase and bovine serum albumin (BSA), the adsorption kinetics is studied. Composite substrates are used enabling a separation of the short- and the long-range forces. Though usually neglected, experimental evidence is given for the influence of van der Waals forces on the protein adsorption as revealed by \textit{in situ} ellipsometry. The three proteins were chosen for their different conformational stability in order to investigate the influence of conformational changes on the adsorption kinetics. Monte Carlo simulations are used to develop a model for these experimental results by assuming an internal degree of freedom to represent conformational changes. The simulations also provide data on the distribution of adsorption sites. By \textit{in situ} atomic force microscopy we can also test this distribution experimentally which opens the possibility to e.g. investigate the interactions between adsorbed proteins.
\end{abstract}



\maketitle

\begin{quotation}

Proteins are found to be involved in interactions with solid surfaces in numerous natural events. Due to protein adsorption and subsequent bacterial adherence, biofilms are formed at interfaces between solid substrates and liquids containing biomacromolecules. Biofilm formation as well as protein adsorption are molecular assembly processes occurring at interfaces \cite{nor1986}. Moreover, these processes are  collective phenomena, since surface coverage, adsorption kinetics, conformational arrangement etc. are depending on the nature of the molecular neigh­borhood. The lack of effective control over this process has  become a bottleneck impeding the development of many biotechnologies, examples including biosensors, enzyme immobilization in biocatalysis, antibody attachment in immunoassays, biomaterial development, and tissue engineering. During the past decades, substantial progress has been made in understanding some mechanisms of protein adsorption \cite{den2002, mal2003, mic2003, gray_interaction_2004, kes2005}. Adsorption is the net result of various types of interactions which depend on the nature of the protein and of the substrate, as well as on the surrounding aqueous solution. Several driving forces have been identified for the protein adsorption process, including dehydration of the protein and substrate surface, redistribution of charged groups in the interfacial film, and structural rearrangements in the biomolecules. Regarding the role of substrate characteristics, both surface topography and surface chemistry have been shown to affect protein adsorption, bacteria and cell adhesion \cite{den2002, teu2006}. However, since most topographical variations are accompanied by chemical heterogeneities, separating both effects is difficult and a differentiation between the impacts of short- and long-range forces is one step towards gaining control of biofilm formation and has only recently been probed \cite{qui2008, bel2008}. It is the aim of this paper to compile the results in order to stress that the subsurface composition is a parameter of utmost importance.

The separation of these short- and long-range interactions is possible using composite samples: By variation of the silicon oxide thickness on top of Si wafers, the van der Waals interactions can be tuned \cite{see2001a, see2001b, see2005, hub2005}, as described in section \ref{waferset}. The adsorption kinetics of BSA and $\alpha$-amylase are sensitive to these variations of the long-range potential. Note that the van der Waals interaction depends on the geometry of the interacting objects \cite{isr1992}: For two atoms, the van der Waals potential decreases like $\bm{1/r^{6}}$, where $\bm{r}$ denotes the distance. However, for a spherical particle interacting with a planar surface the decrease is much slower i.e. like $\bm{1/r}$. Hence, the strength of the van der Waals forces cannot be neglected.

Studies concentrating on the kinetic behavior during adsorption, desorption, and exchange processes revealed that there is a certain hierarchy or intermolecular arrangement of proteins in the adsorption layers (e.g. Schmidt et al. \cite{sch1990}). Additionally, reconfiguration or reorientation of the adsorbed protein can take place \cite{wer2002, ish2005}. However, further information is needed to describe the structure of the adsorption layer in detail. A manifold of proteins have actually been detected to undergo conformational changes or reorientations when adsorbed on a variety of hydrophilic, hydrophobic, charged, and uncharged surfaces (e.g. Keselowsky et al. \cite{kes2005}, Daly et al. \cite{dal2003}, Ishiguro et al. \cite{ish2005}, Baujard-Lamotte et al. \cite{bau2008}). These conformational changes and reorientations were interpreted to be a result of mutual interactions between the adsorbed protein molecules (``critical crowding''), as well as interactions between the proteins and between the proteins and the surfaces. Hence, monitoring the conformational change in an adsorbed protein as a function of the adsorption amount and time should be useful for obtaining information about the intermolecular interactions in the adsorption layer.

The direct detection of conformational changes is a challenging task. Most methods are restricted to indirect observations. When monitoring the adsorption kinetics, different kinetic behaviour is found for the 'rigid' lysozyme \cite{nor1992} than for $\bm{\alpha}$-amylase and BSA \cite{qui2008, bel2008}. The latter is considered as a 'soft' protein \cite{fos1977, car1994, nor1992, oli1995}. This difference in the adsorption kinetics is therefore ascribed to dissimilar tendencies to undergo conformational changes due to adsorption. Monte Carlo simulations including an internal degree of freedom to incorporate conformational changes qualitatively reproduce the experimental findings.

Most methods, like ellipsometry or scattering techniques provide information that is obtained by averaging over some $\bm{\mu}$m$\bm{^{2}}$ . Therefore, these methods cannot provide explicit information about single proteins involved in the adsorption process. A technique that overcomes these shortcomings is atomic force microscopy (AFM). By \textit{in situ} AFM in aqueous environment, the spatial distribution of individual adsorbed proteins can be determined \ref{in situ AFM imaging} and can therefore  give new insights into the evolution of protein films. The surface mobility of the single proteins can be tested as well as their tendency to form clusters or to adsorb rather independent from each other, as described in the scenario of random sequential adsorption (RSA) which is presented in section \ref{RSA}.

\end{quotation}

\section{\label{waferset}materials}
The proteins $\alpha$-amylase (product no. 10092), lysozyme (product no. 62971) and bovine serum albumin (product no. A3059) were purchased from Sigma Aldrich, Steinheim, Germany. BSA (66\,kDa)\cite{hir1990} and $\alpha$-amylase (58\,kDa)\cite{hu2005} have about the same size  but different isoelectric points (pI) at pH 4.7 \cite{arn2003} and 6.5 \cite{hu2005}, respectively. For bovine serum albumin (BSA) the ability of showing different conformations is well known \cite{fos1977, car1994, nor1992, oli1995}. Lysozyme has a mass \cite{rot1995} of 14\,kDa and its isoelectric point \cite{mal1978} is at pH 11.

As substrates for ellipsometry measurements, silicon surfaces with different silicon dioxide thicknesses were used: natural (2\,nm, Wacker Siltronic AG, Burghausen, Germany) and thermally grown (192\,nm, Silchem, Freiberg, Germany) dioxide layers, denoted as thin and thick oxide layers in the following. Before usage, the wafers must be cleaned to remove residues from the polishing procedure (mostly hydrocarbons). Therefore, the wafers were immersed into fresh 1:1 H$_{2}$SO$_{4}$(conc.)/H$_{2}$O$_{2}$(30\%) solution for 30 min. Then the acids were removed by rinsing the wafers in hot deionized water, three times 20 min each. All samples were prepared in a clean room environment (class 100), resulting in a water contact angle of 0$^\circ$ directly after cleaning. Nevertheless, upon usage, the silicon wafers showed advancing water contact angles of less than 10$^\circ$ (see Table~\ref{tab:wafer characterisation} for details) and are therefore termed 'hydrophilic'. To gain a second set of wafers with a 'hydrophobic' surface ($\Theta >$ 90$^\circ$), a self-assembled monolayer (SAM) of silane molecules with a CH$_{3}$ tailgroup (octadecyl-trichlorosilane, OTS, Sigma Aldrich, Germany) was applied to the wafers following standard procedures \cite{was1989, brz1994}. The water contact angles on hydrophobized wafers were lager than 110$^\circ$ (Table~\ref{tab:wafer characterisation}) with a hysteresis smaller than 5$^\circ$. Contact angle hysteresis describes the difference between advancing $\Theta_{a}$ and receding $\Theta_{r}$ contact angles. By additionally measuring the contact angles for glycerol and 1-bromonathalene, the surface energies as well as the Lifshitz-van der Waals and the Lewis acid-base components can be determined \cite{myk2003} as listed in Table~\ref{tab:surface energies}.
\begin{table*}[ht]
                \caption{\label{tab:wafer characterisation}Silicon wafer characterization results.}
\begin{ruledtabular}
\begin{tabular}{ccccccc}
                        d (SiO) / nm    & hydro- & rms / nm & $\Theta^{water}_{a}$ & $\Theta^{water}_{r}$ & $\Theta^{glycerol}_{a}$ & $\Theta^{1-bromonaph.}_{a}$\\
                                \hline
                                2.0(1) & philic & 0.09(2) & 5(2)$^\circ$ & complete wetting & 11(3)$^\circ$ & 13(4)$^\circ$\\
                                192(1)& philic & 0.13(3) & 7(2)$^\circ$ & complete wetting & 17(3)$^\circ$ & 15(3)$^\circ$\\
                                2.0(1) & phobic & 0.12(2) & 111(1)$^\circ$ & 107(1)$^\circ$ & 95(2)$^\circ$ & 62(4)$^\circ$\\
                                192(1) & phobic & 0.15(2) & 112(1)$^\circ$ & 108(2)$^\circ$ & 92(2)$^\circ$ & 63(3)$^\circ$\\
\end{tabular}
\end{ruledtabular}
\end{table*}

\begin{figure}
                \includegraphics[width=\linewidth]{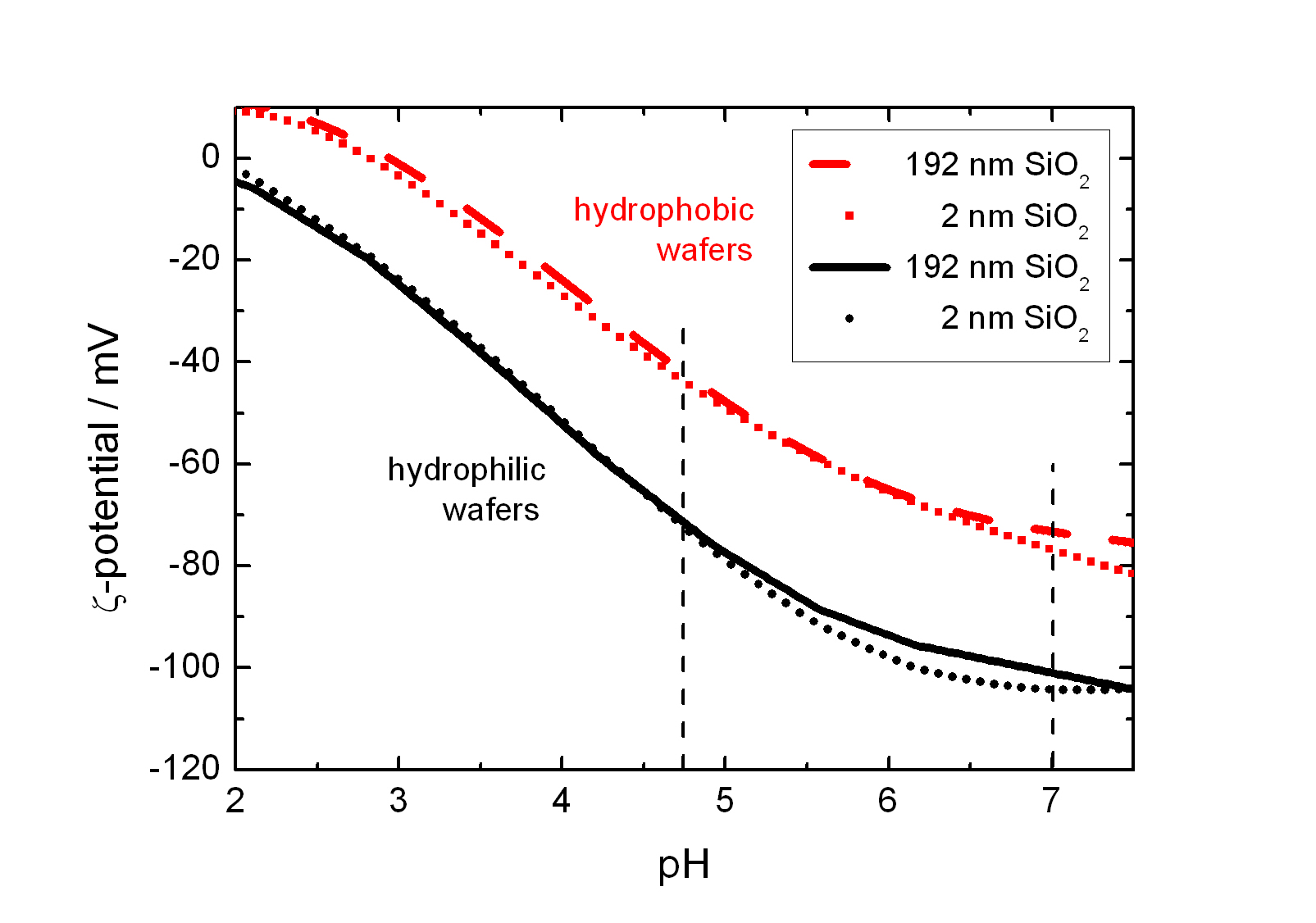}
        \caption{\label{fig:zetapotential}$\zeta$-potential of hydrophobic and hydrophilic silicon wafers with thin and thick oxide thickness as function of pH. For the $\zeta$-potential, differences in oxide layer thickness are irrelevant. Note that adsorption experiments in this study have been performed at pH 7.0 and 4.75 (vertical dashed lines).}
\end{figure}

\begin{table*}[ht]
                \caption{\label{tab:surface energies}Surface energy $\gamma$ and their Lifshitz-van der Waals $\gamma^{LW}$ and Lewis acid-base $\gamma^{AB}$ components of the substrates, as determined by contact angle measurements, the results of which are listed in Table~\ref{tab:wafer characterisation}.}
\begin{ruledtabular}
\begin{tabular}{ccccc}
                        d (SiO) / nm    & hydro- & $\gamma$ / (mJ/m²) & $\gamma^{LW}$ / (mJ/m²) & $\gamma^{AB}$ / (mJ/m²) \\
                                \hline
                                2.0(1) & philic & 64.2 & 43.5 & 20.7\\
                                192(1) & philic & 63.2 & 43.1 & 20.1\\
                                2.0(1) & phobic & 24.1 & 24.1 & 0.0\\
                                192(1) & phobic & 23.6 & 23.6 & 0.0\\
\end{tabular}
\end{ruledtabular}
\end{table*}
Zeta-($\zeta$)-potential measurements \cite{zim2001} on wafers with thin and thick oxide layers with and without the OTS monolayer were performed for pH values ranging from 2.0 to 7.5 (see Fig. \ref{fig:zetapotential}). Obviously, the hydrophilic surfaces carry a higher negative charge than the hydrophobic wafers. The oxide layer thickness, however, does not play any role and zeta potential values are identical within the experimental error.

%

The thicknesses of the silicon oxide and OTS layer were determined by ellipsometry, revealing an oxide thickness of 2\,nm for the thin and 192\,nm for the thick oxide layers. The thickness of the OTS-SAM is found to be 2.6\,nm. The surface roughness of the four wafer types was determined by atomic force microscopy (AFM). The measured root mean square (rms) roughness of an (1 $\mu$m)$^2$ scan area was below 0.2\,nm for all wafer types (see Table~\ref{tab:wafer characterisation} for details).

Keeping all other parameters (pH, ionic strength, temperature and protein concentration) constant, the four wafer types allow for the separation of effects due to the long-range and the short-range part of the surface potential \cite{see2001a, see2001b, see2005, hub2005}: Hydrophobization by the OTS-SAM alters the short-range part as can be seen in the contact angle and the $\zeta$-potential. For a substantial alteration of the long-ranged van der Waals potential, the OTS-SAM is too thin. This holds for all molecular-sized layers. The variation of the oxide layer thickness, however, does not have an influence on the short-range potential, since the surface chemistry is identical. Hence contact angle as well as $\zeta$-potential measurements show no difference on thin or thick oxide layers. Vice versa, choosing different subsurface wafer compositions (in our case thin and thick oxide layers) enables an independent variation of the long-range part, i. e. the van der Waals interaction to probe its influence on the formation of protein films.

The above described four types of Si wafers were used for probing protein film formation by ellipsometry. For the \textit{in situ} AFM experiments shown in the following, mica was chosen as a substrate. Mica is a sheet silicate that can easily be cleaved. When freshly cleaved, mica exhibits an atomically smooth hydrophilic surface. Therefore, mica is frequently used for protein adsorption studies via AFM \cite{kim2002,sil2005, che2008, oue2002, ber1998, mar2003}.

\section{adsorption kinetics}

\subsection{Adsorption studies by ellipsometry}

Ellipsometry\cite{cuy1983, voe2004} was used to determine the adsorbed amount of proteins by \textit{in situ} monitoring the adsorption process. The single wavelength ellipsometer (EP$^{3}$, Nanofilm, G\"ottingen, Germany) was operated in PCSA (Polarisator-Compensator-Sample-Analyzer) configuration at a wavelength of 532\,nm. The ellipsometric angles $\Psi$ and $\Delta$ were recorded via the nulling ellipsometry principle with a sampling rate of 1.5 to 6 per minute. This rate was sufficient to monitor the formation of protein layers \textit{in situ}.

To determine the physical properties of the reflecting surface from $\Psi$ and $\Delta$, a model has to be applied \cite{cuy1983}. For single wavelength measurements, one has to assume the layers to have a constant height to be able to determine the refractive index and thickness of all layers. For layers with a thickness below 5\,nm, it is not possible to distinguish between a change in refractive index and in film thickness \cite{azz1977}. Therefore, de Feijter's method is applied to determine the adsorbed amount $\Gamma$ as a function of the refractive index $n_{f}$ of the protein film with a fixed film thickness $d_{f}$
\begin{equation}
\Gamma = d_{f}\frac{n_{f}-n_{a}}{dn/dc}
\end{equation}
where $n_{a}$ denotes the refractive index of the ambient and $dn/dc$ the increment of the refractive index of the solution due to the increase of molecule concentration. The refractive index $n(c)$ is assumed to be a linear function\cite{fei1978, bal1998} with a fixed gradient of 0.183\,cm$^{3}$/g.

The measurements were carried out in a temperature controlled closed fluid cell made of Teflon. A connection to a flow system enables a continuous exchange of fluid with constant flow rates and the injection of protein solution via a sample injector (Rheodyne Manual Sample Injector).

The proteins were soluted in a 10 mM phosphate buffer solution (pH 7.0, ionic strength 20\,mM). The fluid system was filled with the same buffer and the buffer was run through the fluid cell at a constant flow rate. After reaching a thermal equilibrium (either at 37$^{\circ}\mathrm{C}$ or at room temperature) associated with a constant baseline, the adsorption measurement was started by injecting the protein.

\subsection{Results and discussion}

\begin{figure*}
                \includegraphics[width=\linewidth]{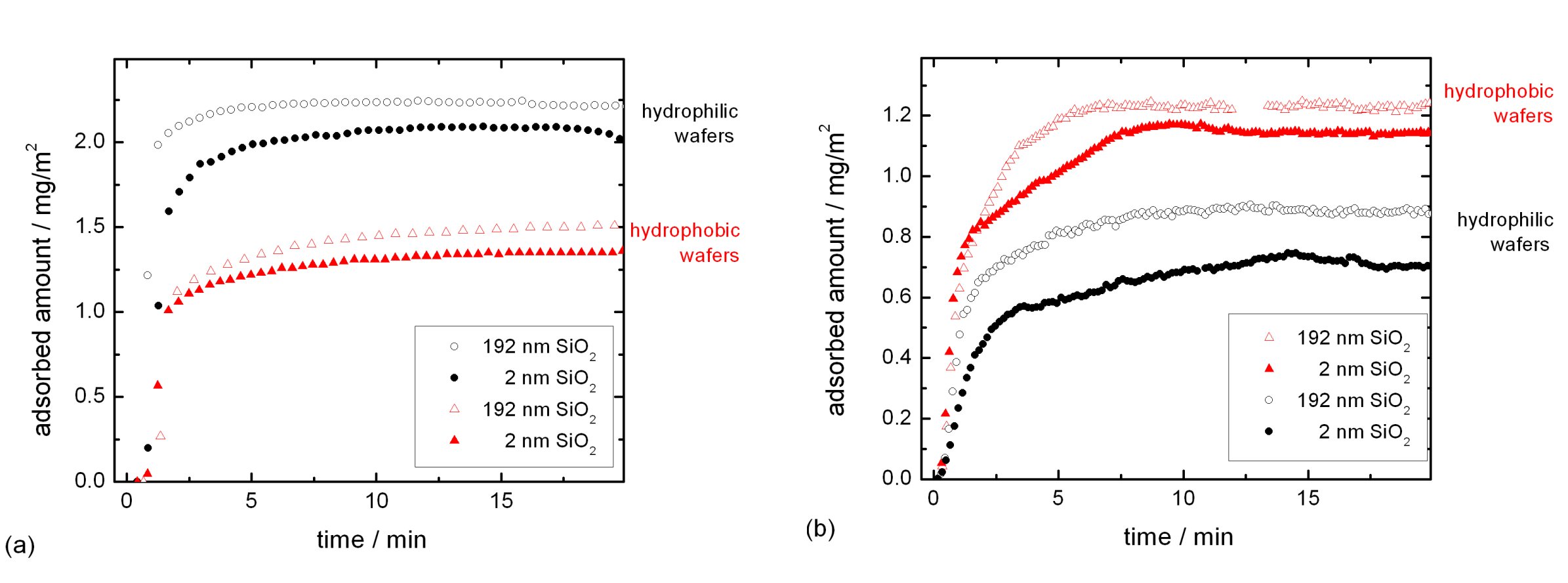}
        \caption{\label{fig:lysozymeandamylase}Adsorption kinetics of (a) lysozyme and (b) amylase on the four different
types of surfaces.}
\end{figure*}

The adsorption kinetics are shown in Fig. \ref{fig:lysozymeandamylase}. They display a typical feature of adsorption kinetics: after a fast initial adsorption, the adsorption rate decreases until a plateau value of adsorbed amount is reached. For lysozyme (Fig. 2 (a)), the decrease of adsorption rate is continuous. This Langmuir-like adsorption kinetics is well described in the literature \cite{mal2003}. The absolute adsorbed amount depends mostly on the chemistry of the surface. The adsorbed amount of lysozyme is higher on hydrophilic samples than on hydrophobic ones. This can be explained by considering the positive net charge\cite{rot1995} of lysozyme (+\,8\,e) at pH 7 as well as the negative charge of the silicon wafers. The hydrophilic wafers (-103\,mV) carry a higher negative charge than the hydrophobic ones (-75\,mV), see Fig. \ref{fig:zetapotential}. Thus, a stronger attractive Coulomb interaction promotes the adsorption on hydrophilic wafers. For lysozyme, we observe no influence of the thickness of the oxide layer on the adsorption kinetics.

For $\alpha$-amylase (Fig. 2(b)), the adsorption on thick oxide wafers also exhibits a Langmuir-like behavior \cite{qui2008, bel2008}.  On thin silicon oxide, however, the transition between fast initial adsorption and saturation is not continuous. A regime of constant adsorption rate separates the regime of initial adsorption from the saturation. The final adsorbed amount depends only weakly on the oxide thickness but on the hydrophobicity of the surface. In the case of $\alpha$-amylase, the adsorbed amount is higher on hydrophobic wafers than on hydrophilic ones due to the hydrophobic effect \cite{tan1978}. The higher attraction towards the surface originates from minimizing the contact of water with hydrophobic side groups of the protein and with the hydrophobic surface. As the pH of the solution is close to the isoelectric point of $\alpha$-amylase (pI 6.9), the protein carries no net charge. Therefore, the Coulomb interaction just plays a minor role for adsorption affinity.

The differences in adsorption behavior between lysozme and $\alpha$-amylase are believed to arise from different conformational stabilities \cite{bel2008}. Changes of conformation due to adsorption are frequently described in the literature \cite{nor1992}. As compared to the small and compact lysozyme, $\alpha$-amylase can be ascribed as a `soft' protein \cite{mal2003}. Therefore, major conformational changes are more likely for $\alpha$-amylase than for lysozyme.

\begin{figure*}
                \includegraphics[width=\linewidth]{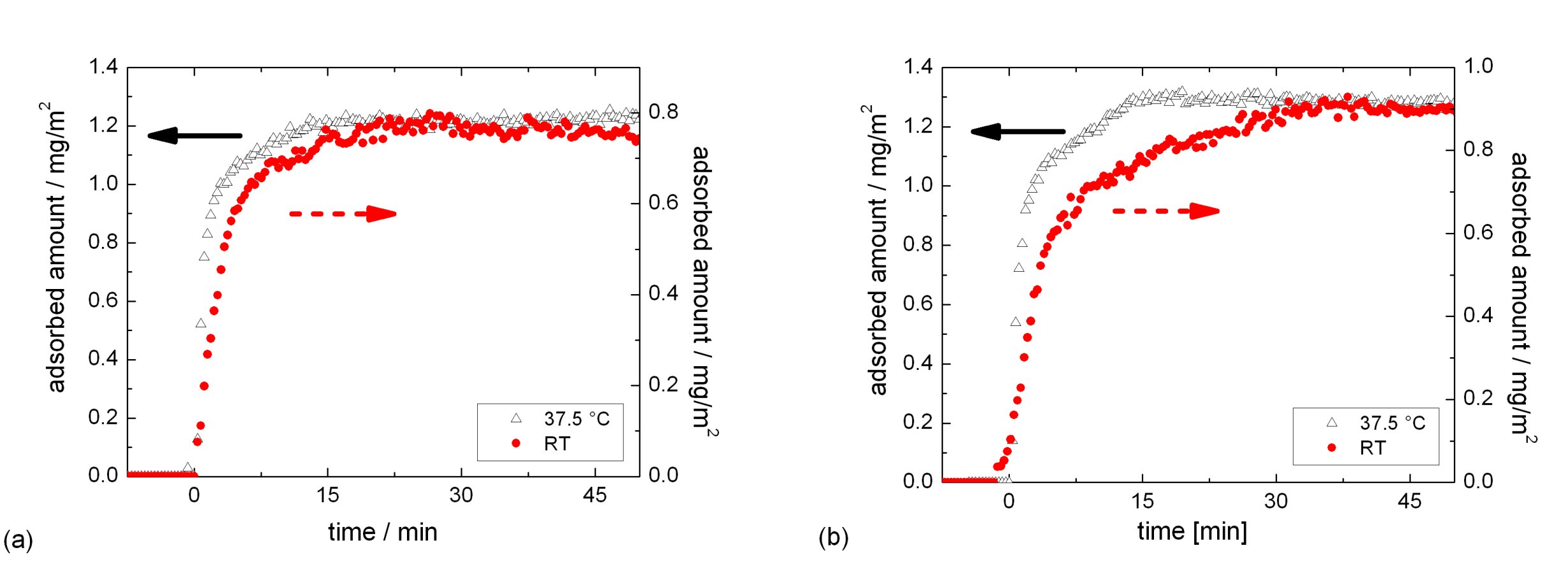}
        \caption{\label{fig:BSAtemperature}BSA adsorption kinetics on hydrophobized silicon wafers with (a) thick and (b) thin oxide layers. Measurements at room temperature (RT) and 37.5$^{\circ}\mathrm{C}$ are shown. A regime of constant adsorption rate can only be observed on wafers with thin oxide layers and the intermediate linear regime is elongated at room temperature.}
\end{figure*}

The third protein in this study, BSA, is known for its ability to change conformation \cite{fos1977, car1994, nor1992, oli1995} e.g. due to changes of the pH of the surrounding solution. As shown in Fig. \ref{fig:BSAtemperature}, BSA exhibits a similar adsorption kinetics as $\alpha$-amylase \cite{bel2008}. On wafers with thin silicon oxide layer (here only experiments on hydrophobic wafers are shown) a regime of constant adsorption rate is observed. When decreasing the temperature from physiological 37$^{\circ}\mathrm{C}$ to room temperature, the regime of constant adsorption rate is elongated. The initial adsorption, however, is only slightly slowed down. Although BSA carries a negative net charge at pH 7 (in contrast to $\alpha$-amylase), it displays the same two types of adsorption kinetics as $\alpha$-amylase. Repeating the experiments with acetate buffer at the pI of BSA reveals the same results. Therefore, this effect cannot be explained by charge effects and Coulomb interactions. The appearance of a linear regime must rather depend on the variation of the long-range part of the interaction potential i.e. the van der Waals interaction as well as the conformational stability of the protein.

At this point it has to be stressed that numerous experimental studies focussing on thin film formation or dynamics usually do not give details on the subsurface composition of the samples in the study. The subsurface composition therefore is one to date mostly uncontrolled sample property. It is very likely that many diverse results about biofilm formation of different groups world-wide may be traced back to using different composite samples (still exhibiting the identical surface chemistry).
We therefore pledge for a detailed characterization of samples, including e.g. a possible layering. For the widely used Si wafer, for instance, it is inevitable to characterize its oxide layer\footnote{Being precise, not only oxide layer thickness shall be characterized, also the refractive index, since depending on the production process, different porosities of amorphous Si dioxide layers can be gained.}.

Since an experimental study of possible conformational changes of individual proteins during the adsorption process is too challenging to date, colloidal Monte Carlo (MC) simulations were launched using an effective particle model. Here, the sensitivity of the adsorption kinetics to conformational (or even simpler: geometrical) changes of adsorbing objects shall be probed and the results shall then be compared to the experimental findings in order to be able to propose a suitable model. At first sight, a colloidal approach seems too rough to give reliable results for protein adsorption. Yet after checking possible alternatives - which will be done in the following - it shall become clear that -- to match the time scale of the experiment, to be able to include conformational changes and to use to date computer power -- colloidal MC simulations are the method of choice.

\section{Monte Carlo Approach}
The numerical investigation of biofilm adsorption is a challenging task since the
number of degrees of freedom is huge if the detailed structure of the proteins
is taken into account. In addition to this, proteins undergo structural changes
at the surfaces which implies that the relaxation time for the adsorption of a
single protein at the surface is out of scope for present simulation approaches.
Therefore, a drastic reduction of the degrees of freedom is mandatory in order to
simulate the adsorption of a protein film. In this section, we will briefly discuss
the scope of different numerical approaches to protein dynamics.

The most fundamental methods are based on explicit quantum mechanical calculations.
Incorporating the distribution of electrons they are able to predict the forming
of atomic bonds, which is indispensable for studying chemical reactions of
proteins (see ref.\cite{marx_proton_2006} for an overview).
Ab initio methods approximately solve Schr\"odinger's equation. Being able to
provide very accurate results the calculations are very time consuming. To date,
only a two-digit number of atoms can be simulated using this ab initio
approach\cite{latour_molecular_2008, rimola_ab_2009}.  Therefore, procedures have
been developed\cite{nikitina_mixed_2006, nikitina_semiempirical_2004, wada_quantum_2005,yu_refinement_2005,friesner_ab_2005}
simplifying the unessential parts of the system using a semiempirical approach
while describing the important parts using quantum mechanical methods. In these
cases, the solvent is partly implicit.

Much larger systems can be considered by using fragment methods. Retaining the
accuracy\cite{gordon_accurate_2009}, these divide the whole system into several
subunits which can be processed individually enabling the use of parallel
computers. Differing in the underlying method, several fragment methods are
known, e.g. FMO (fragment molecular orbital) method
\cite{fedorov_extendingpower_2007, li_energy_2010}, SFM (systematic
fragmentation method)
\cite{deev_approximate_2005,collins_accuracy_2006,collins_molecular_2007} and
EFP (effective fragment potential)
method\cite{gordon_effective_2001,minikis_accurate_2001}. Fragmentation methods
have the advantage that they are nearly fully quantum mechanical in nature, with
classical approximations often used for long-range interactions. They can be
used for the simulation of small proteins:
Ishida analyzed\cite{ishida_probing_2008} the role of a polar protein
environment in a special enzymatic process using FMO. Ishikawa et al
determined\cite{ishikawa_theoretical_2009} the molecular interactions between
the prion protein and GN8 (a potential curative agent for prion diseases) based
on the fragment molecular orbital method, too. Due to the limitation of the
system size and time scale, however, they can't be used in order to study protein
adsorption involving conformational changes.

To overcome those limitations, quantum effects are often neglected.
This is also the case for so-called full atomistic molecular dynamic simulations,
where atoms are the smallest building blocks. By using these classical
approaches, the simulation of peptides and small proteins becomes feasible.
This includes forecasts of structural changes of single proteins and forecasts
of possible transition paths between two states
\cite{bolhuis_transition_2002,kim_elastic_2002}. It also includes the
simulation of changes of conformation, such as simpler ones that involve only
loop motions. Those take place on a nanosecond time scale, while large changes
that involve the rearrangement of domains may occur on the microsecond to
millisecond time scale \cite{kern_role_2003}. These larger changes are still not
accessible using an atomistic level of detail.

The time evolution of a system can be obtained via Molecular Dynamics, Brownian
Dynamics or local Monte Carlo simulations, but the latter two have a limited ability
of simulating the solvent, as they can only do it implicitly. Often this is done
in MD, too, as means of speeding up a simulation, retaining the
atomistic precision of the protein model.
Whether this is reasonable for incidents like protein folding or adsorption is
an open question\cite{freddolino_ten-microsecond_2008}. Some results suggest
that it is good enough\cite{voelz_molecular_2010}, at least for folding.
Another comparison of the influence of the level of detail of the solvent on
the adsorption of peptides on hydrophobic surfaces is done by
Sun and coworkers\cite{sun_comparison_2006,sun_comparison_2007} who find, that implicit
solvent models yield very different results as compared to the explicit solvent model
which was used as a reference. Also the different implicit solvent
models predict different binding energies with varying level of agreement\cite{sun_comparison_2006}.

Further simplification for speeding up the simulations is done by combining multiple
atoms to a single pseudo atom.
This coarsed system needs to be described using modified force fields or
interaction energies, respectively. For a fixed system size, the accessible time scales are
two to three orders of magnitude higher than for a full all-atom
description\cite{monticelli_martini_2008}, so in principle they allow the
simulation of large proteins. So far, however, the currently available generic
force fields don't support structural changes in the tertiary
structure\cite{monticelli_martini_2008}. Therefore, reliable simulations can be obtained
\emph{only} if the structure of the protein is maintained. An overview of those
so called united atom or. coarse-graining methods is given in ref.
\cite{tozzini_coarse-grained_2005,bereau_generic_2009}.

Classical MD methods are based on the integration of Newton's equations.
The numerical effort is very high, typical time scales are from 10\,ns
\cite{karplus_molecular_2002} up to some  10\,$\mu$s
\cite{freddolino_ten-microsecond_2008,voelz_molecular_2010}, depending on the
system size, the hardware being used  and accuracy (ex- or implicit water) of
the calculations. Raffaini et al for example were able to carry out simulations of
the adsorption of single protein domains with $\alpha$-
\cite{raffaini_simulation_2003} and $\beta$-sheets \cite{raffaini_molecular_2004}.
Later on they analyzed\cite{raffaini_sequential_2007} the successive adsorption
of protein  fragments. The adsorption of a complete but small protein, lysozyme,
including its conformational changes was simulated by Kubiak and Mulheran
\cite{kubiak_molecular_2009}. Details of the large scale MD
simulations can be found in  ref.\cite{klepeis_long-timescale_2009}.

Omitting the time evolution completely, greatly simplifying the solvent and the
intermolecular interactions to the point of completely stiff proteins, interaction
energies used to be calculated since two decades \cite{lu_protein_1990}. There
are more recent publications using this
approach\cite{camacho_protein_2001,hsu_preferred_2008} proposing docking paths
and preferred orientations, but conformational changes cannot be predicted by
this procedure.

The range of applications for full atomistic as well as coarse-grained MD
simulations indicate that these approaches cannot be used in order to
simulate the adsorption of a full biofilm.

Speeding up the simulations by describing the interaction with the solvent
molecules stochastically using the Langevin-equation results in a Brownian
Dynamics simulation. Random forces with zero mean value mimic collisions of
solvent molecules with the beads of a protein.
Ravichandran et al\cite{ravichandran_brownian_2001}  were able to show the
possibility of adsorption of proteins with a positive net charge to a positively
charged surfaces due to inhomogeneous charge
distribution. Since then, however, the number of
publications using Brownian Dynamics at atomistic level of detail is declined
in favor of Molecular Dynamics.

Monte Carlo simulations provide another stochastic ansatz. They can be used both
for time evolution of a system (when using local moves solely and detailed
balance to reach a stationary state) or transition paths between two states.
Simulations of coarse-grained lattice proteins were published some time ago:
Anderson et al\cite{anderson_dynamic_2000} studied the adsorption of a model
protein to an oil/water interface and Castells and
coworkers\cite{castells_surface-induced_2002} analyzed surface induced
conformational changes. Being more widespread and potentially more accurate,
today's researchers tend to prefer the usage of MD simulations as a workhorse
when generating the time
evolution at atomistic level. Nevertheless, due to their flexibility, Monte Carlo
simulations offer an alternative method: Knowing an initial and a final state of
a protein, it is efficiently possible to sample the transition path between those
two states\cite{bolhuis_transition_2002,yi_qin_gao_thermodynamics_2008,escobedo_transition_2009}.
Sadly, the position of atoms in a protein/surface complex can't be extracted
experimentally\cite{makrodimitris_structure_2007, gray_interaction_2004}.
Not knowing the adsorbed state beforehand, this method cannot be used for
protein adsorption.

A more detailed review of the methods used for the simulation of conformal changes
can be found in ref.\cite{van_der_vaart_simulation_2006}.

\subsection{\label{RSA}Collodial approaches}

Coarsening the system under consideration even more, colloidal length scales  of
 a few nm are reached. This results in a \emph{substantial} increase of both
spacial and temporal time scale, the latter up to the order of seconds to hours,
depending on the level of complexity. In this context, the solvent is only
implicitly taken into account and proteins typically are modelled as
particles of high symmetry, e.g. spheres, allowing many thousands of proteins to
be considered. The downside is the neglect of all their structural
details including the distribution of charges and other details of the
interactions which are relevant for very short distances. Yet, this is a
necessary compromise in order to reach time scales that are experimentally
relevant for the adsorption of a protein-biofilm.

A simple but very efficient approach to protein adsorption is the so-called
Random Sequential Adsorption (RSA), which typically is a 2D model describing the
parking problem omitting the particle's trajectory to the surface.
In the simplest case, at every time step a particle
tries to adsorb irreversibly to a random site by chance. The probability is
determined  using the Metropolis-Hastings algorithm in conjunction with a
hard core potential. So the surface coverage only depends on the volume exclusion.
The jamming coverage of this problem is described by
Hinrichsen et al\cite{hinrichsen_geometry_1986} and the maximum theoretical coverage
by Toth\cite{toth_densest_1983}.
There are several enhancements made to this simple model. The adsorption and
desorption of particles in combination with a non-spherical geometry and a
change of conformation was considered by van Tassel et al\cite{tassel_kinetic_1997}.
Also more complex potentials
\cite{talbot_car_2000, adamczyk_unoriented_1997, oberholzer_2-d_1997}, e.g. for
 soft-RSA\cite{semmler_diffusional_1998,adamczyk_structure_1990} are taken into
consideration.
Furthermore, the effects of quasi-3D models using convex particles were studied
by van Tassel and coworkers\cite{tassel_irreversible_1994, tassel_kinetic_1997}.
Lavalle et al\cite{lavalle_extended_1999} analyzed an RSA variant,
in which adsorbed particles are immobilized not before a certain amount of time,
allowing them to find a more suitable spot minimizing the free energy of the adsorbed particles.
Finally, there is a more recent attempt to model the reversible adsorption of
binary mixtures\cite{lonarevi_reversible_2007} resulting in a simple formula for
the prediction of steady state coverage of mixtures depending on the steady
state coverage of the pure components.
In any case, however, the comparison to experiments is limited, as fewer experimental
parameters are realizable and the particle's trajectory to the surface is not
accounted for.

On  the colloidal level, often Brownian Dynamics simulations are used. In this
case, interactions are typically within the DLVO
framework\cite{miyahara_adsorption_2004, ravichandran_mobility_2000, oberholzer_grand_1997}
including Coulomb interactions in a double layer, van der Waals interactions and
additionally steric repulsion. Those publications analyze adsorption phenomena in general,
in the latter, the parameters where chosen to mimic the adsorption of small globular
proteins such as lysozyme. Spontaneous ordering depending on the ionic strength of
the solvent is observed, too\cite{miyahara_adsorption_2004, watanabe_dynamics_2005}.

\subsection{Simulation of a biofilm using colloidal models with internal degrees of freedom}
\begin{figure}[ht]
  \includegraphics[width=0.7\linewidth]{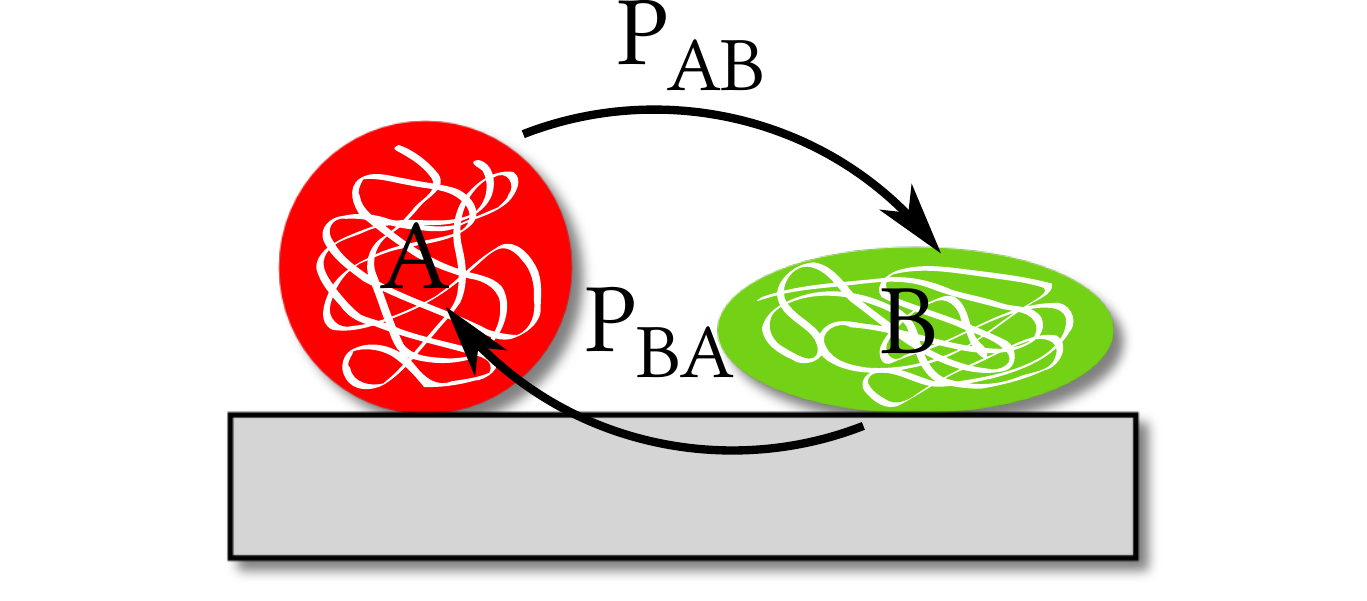}
   \caption{\label{fig:confchange} Being adsorbed to a surface, particles may
           change their conformation by chance from a compact to an expanded
           state and vice versa using probabilities given by the
           Metropolis-Hastings algorithm.}
\end{figure}

\begin{figure*}
  \includegraphics[width=0.8\linewidth]{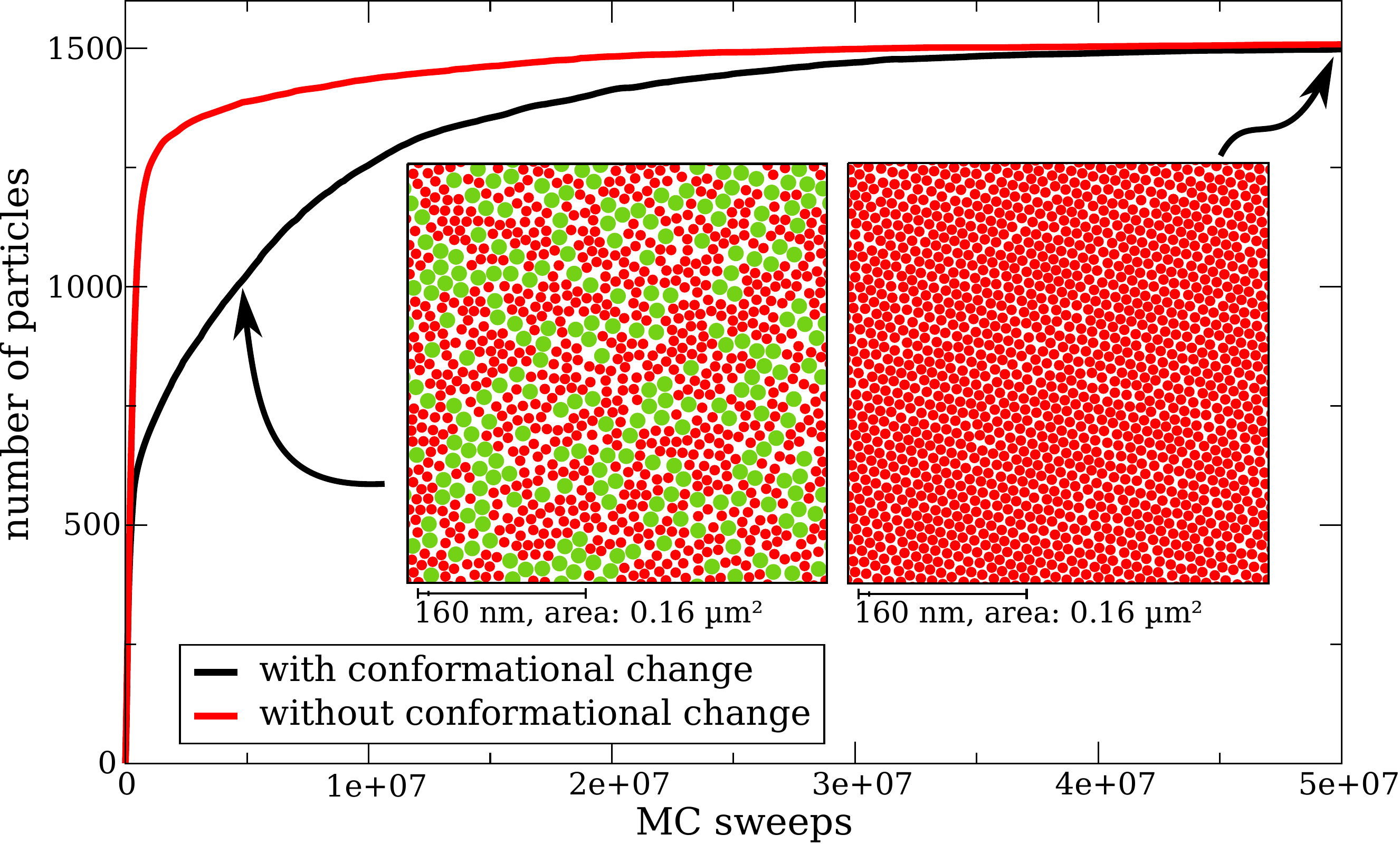}
  \caption{\label{fig:kinetics}The adsorption kinetics of simulations with
           (black) and without (red) change of conformation. For lower surface
           coverages, the expanded state is more stable than the compact one.
           Packing more particles to the surface, the contrary occurs.}
\end{figure*}

Using local Monte Carlo simulations, it is easily possible to include internal
degrees of freedom into the model. We use such an internal degree of freedom to
model a reversible change of conformation of a protein from a compact to an
expanded state on adsorption to the surface as shown in Fig. \ref{fig:confchange}.
The expanded state optimizes the surface interaction at the expense of the
particle/particle interaction whereas the compact state minimizes the covered
surface. Using this particle model we observe \cite{bel2008,qui2008} a
considerable deviation from simple, Langmuir-like adsorption kinetics
(cf. Fig. \ref{fig:kinetics}).

\begin{figure}
    \includegraphics[width=0.6\linewidth]{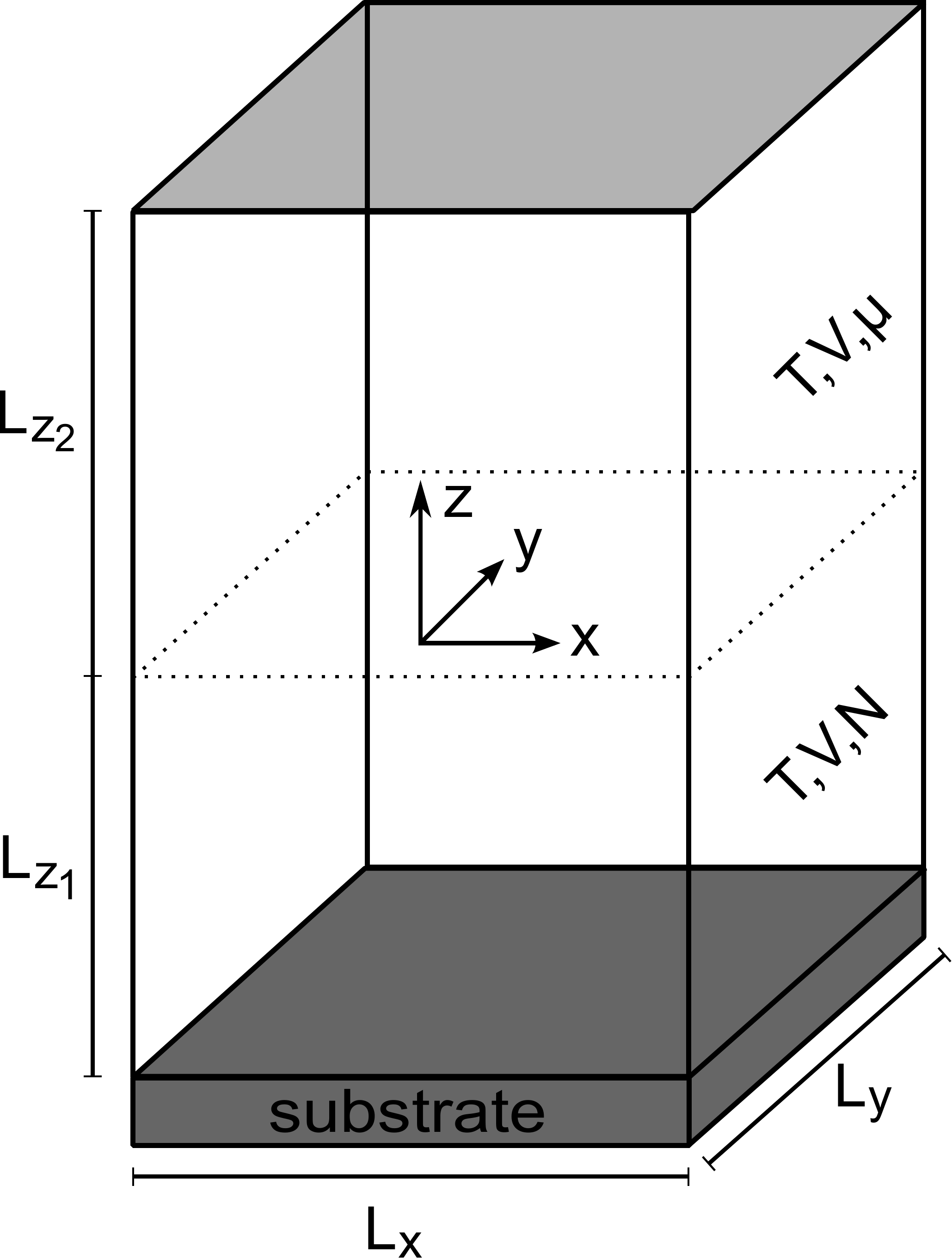}
    \caption{\label{fig:simbox}Simulation box}
\end{figure}

\begin{figure}
   \includegraphics[width=0.98\linewidth]{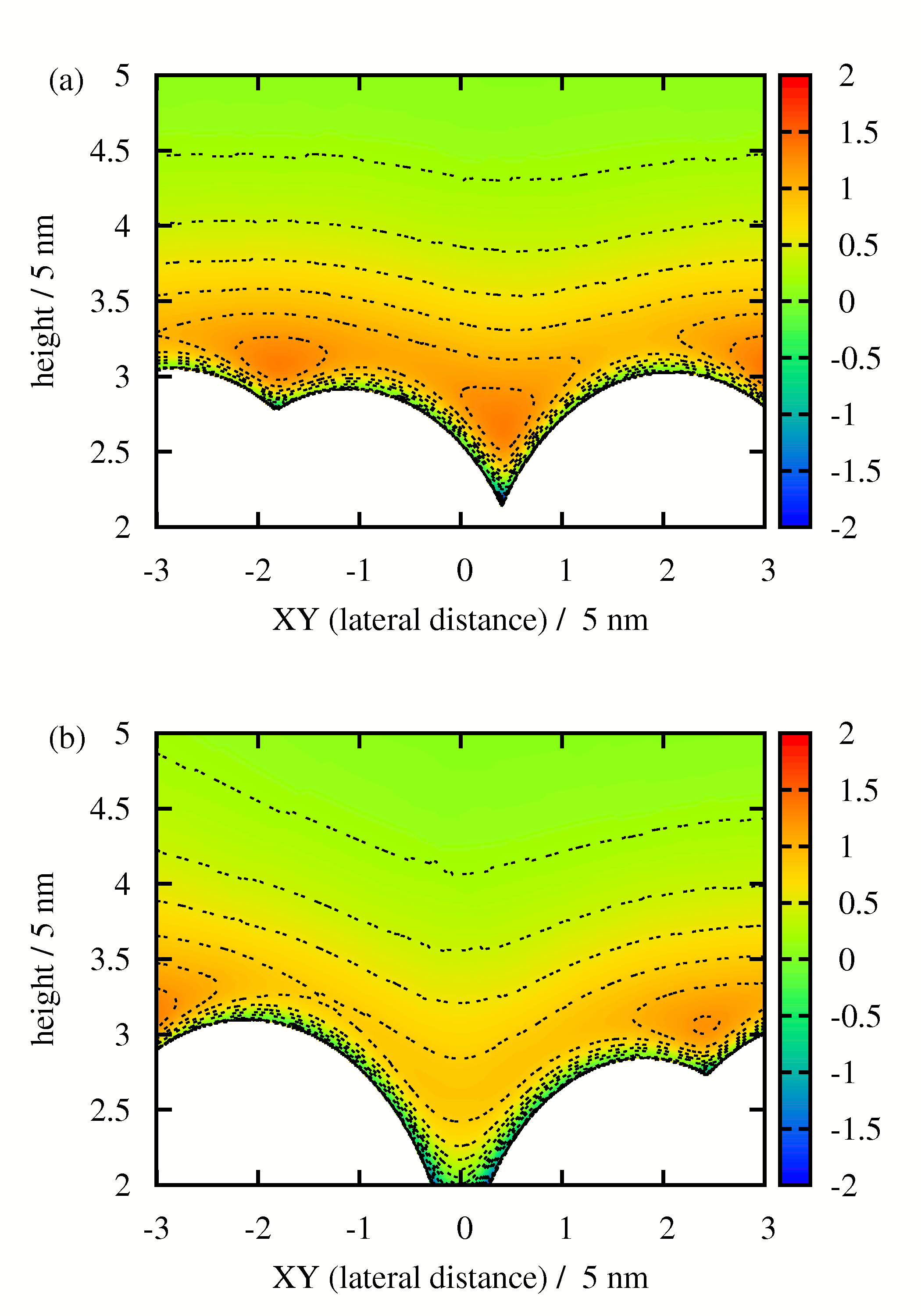}
   \caption{\label{fig:potential}The energy landscape of a particle approaching the
            surface for two different points in time. The isolines mark 0.2
            steps, the white area has infinite potential energy. }
\end{figure}

In this specific model, DLVO interactions where utilized and adjusted using
experimentally known parameters to mimic amylase adsorbing to silicon wafers.
However, due to the simplicity of the protein model many results are at least
on qualitative level valid for other types of proteins as well.

The size of the simulation box is tiny compared to the volume of the experimental
fluid cell. Applying a canonical ensemble for the whole simulation box,
the drift of the protein concentration in the solvent would be largely
overestimated. Therefore it was necessary to use two subvolumes (cf. Fig. \ref{fig:simbox}).
The upper part was to mimic an infinite reservoir of particles using a grand
canonical ensemble, whereas the lower part being in contact with the surface had
to be simulated using a constant number of particles, which was only changed by
particles entering or leaving the box by means of diffusion.

\begin{figure}
   \includegraphics[height=\linewidth,angle=-90]{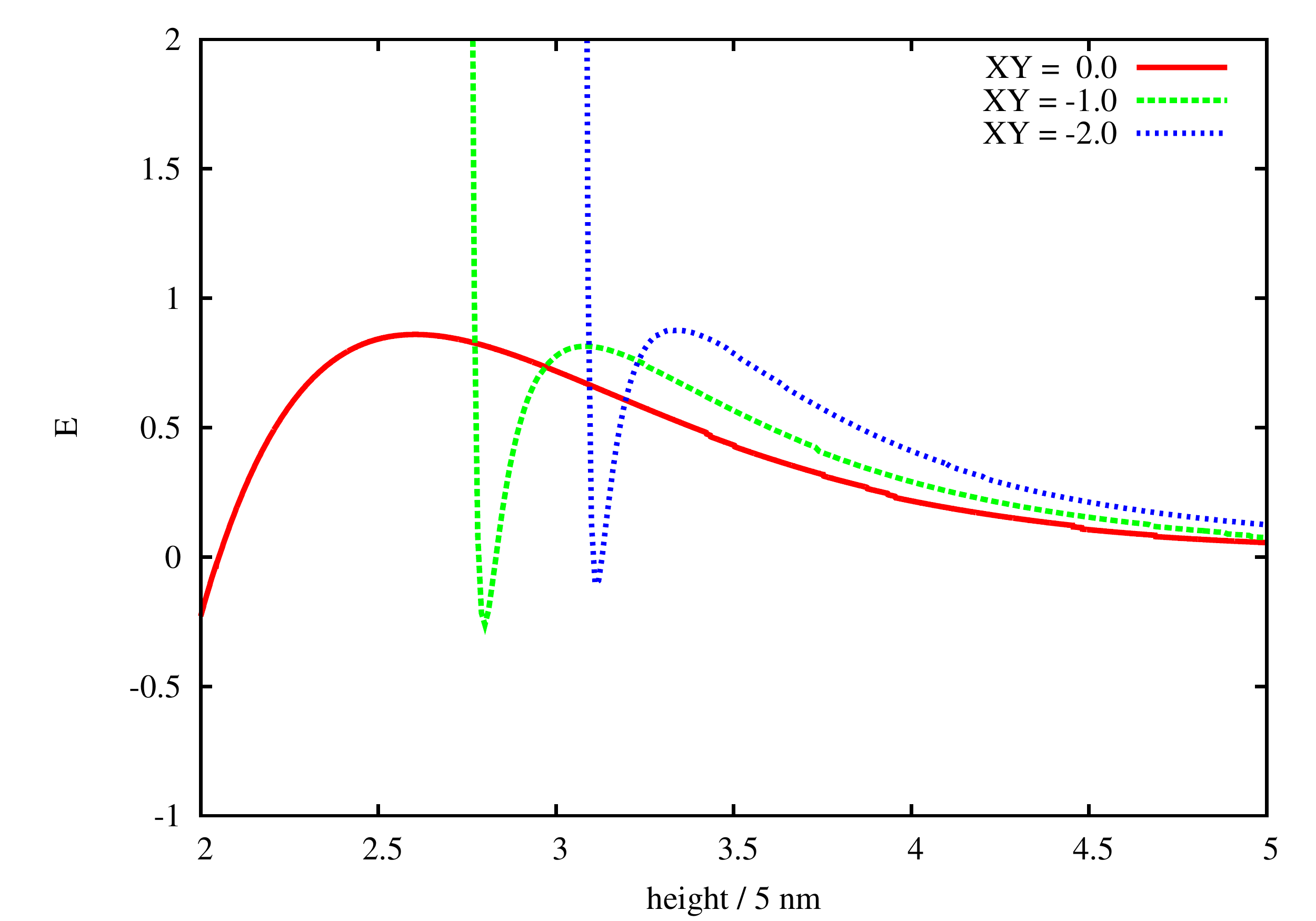}
   \caption{\label{fig:potential-cut} The energy landscape of a particle
           approaching the surface. This is a plot of the potential energy for
           different lateral distances as shown in Fig. \protect\ref{fig:potential} b).}
\end{figure}

Particles adsorbing to the surface are directed to their adsorption site by the
interaction with other particles already adsorbed. This is a major difference
compared to RSA, where possible adsorption sites are chosen by chance.
Fig. \ref{fig:potential} shows two snapshots of the energy landscape a
particle moves in when approaching the surface. In the second plot, there is enough
free space, so that a particle can adsorb to the surface. As a consequence, the
energy barrier evoked by the particles already adsorbed is lower than in the first plot.
Passing this barrier, a particle is trapped in an attractive potential,
as illustrated in Fig. \ref{fig:potential-cut}. Therefore it has to following the
iso line of the minimum and thus, the particle is directed to the adsorption site.
Actually Fig.\ref{fig:potential} b) evolved from \ref{fig:potential} a):
Adsorbing to the surface, a particle moved it's neighbours a bit, so that it could
fit to the adsorption site. The particle was removed again for the plot to
ensure a better comparison of the energy landscape.
Typically the comparison with the experiment is achived via the measured kinetics.
It is suggested by experimental results that the deviation from Langmuir kinetics
is due to long-ranged interactions with the silicon substrate below the oxide layer.
Additionally, simulations indicate a similar deviation caused by
conformational changes which are induced by long-ranged van der Waals
interactions. So, combining experimental and theoretical results, we suppose that
those van der Waals-induced conformational changes are the cause for the
experimentally observed kinetics.
Yet, as the same theoretical curves could possibly be generated in different ways,
comparing the distribution of the adsorbed particles in both simulation and
experiment would be desirable. Using ellipsometry, their positions are not
available. Therefore, statistical properties like the pair correlation or Euler
characteristics cannot be extracted from these experimental measurements.
To overcome this deficiency \textit{in situ} AFM measurements can be performed,
as presented in the next section.

\section{\label{in situ AFM imaging}In situ AFM imaging}

Atomic force microscopy (AFM) was used to image the surface topography of the substrate as well as the adsorbed proteins \textit{in situ}. For the first experiments, mica was used as substrate due to its simple handling and its smooth surface. The measurements were performed in TappingMode\textsuperscript{\texttrademark} to minimize the influence of the scanning tip on the proteins. A triangular cantilever (OTR 8 or SNL, Veeco, Santa Barbara) was oscillated at a frequency of about 9 kHz. The scanning rate was between 0.5 to 1.5 scan lines per second depending on the scanning size (ranging from (0.5\,$\mu$m)$^{2}$ to (3\,$\mu$m)$^{2}$). The lateral resolution of the scans was either 256 or 512 pixels per line. The measurements were carried out in a closed fluid cell (Model MTFML, TappingModeTMFluid Cell, Veeco, Santa Barbara) using a MultiMode (Nanoscope III, Veeco, Santa Barbara). The cell was connected to a flow system (same set-up as in ellipsometry measurements) to allow for the exchange of buffer and protein solution. Experiments were performed at room temperature. Mica being frequently used for protein adsorption studies via AFM \cite{kim2002, sil2005, che2008, oue2002, ber1998,mar2003} was chosen as a substrate.

The mica was mounted in the fluid cell directly after cleavage and consequently exposed to acetate buffer of pH 4.75 at a concentration of 10 mM and ionic strength of 8 mM. A control scan of the mica surface was taken before exposing the surface to protein solution. At intervals of 5 to 10 minutes, the flow was stopped and topographic scans of the surface were taken (see Fig. \ref{fig:BSAonMica1}).
To ensure the absence of scan artifacts, as often reported \cite{lea1992, agn2004, sie1994, epp1993}, the scan area was enlarged several times to check for possible scan damage due to the tip. Under the experimental conditions mentioned above (pH, ionic strength), no influence of scanning on the distribution of the adsorbed proteins was observed. However, increasing the ionic strength of the solution lead to significant disturbance of the protein pattern upon scanning.
\begin{figure*}
                \includegraphics[width=\linewidth]{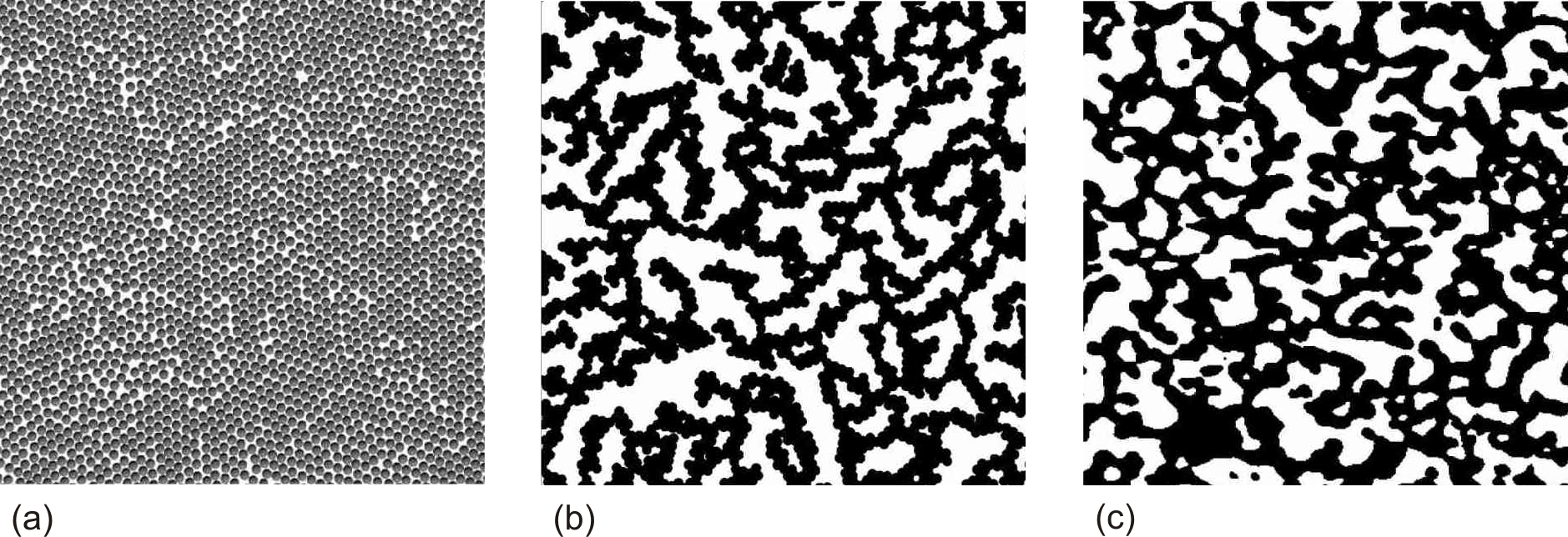}
        \caption{The effect of linker molecules for a stabilization of a protein film: (a) Initial configuration in the MC simulations using the standard form of interactions. (b) Stationary configuration in the computer simulation after adding a short ranged contribution to the particle-particle interactions to mimic the effect of a linker. (c) Experimental view of an amylase film stabilized by glutaraldehyde. For a simple comparison with the simulations, a threshold has been applied to the hight scale such that protein is black and substrate is white.}
        \label{fig:linkermolekule}
\end{figure*}
\begin{figure}
                \includegraphics[width=\linewidth]{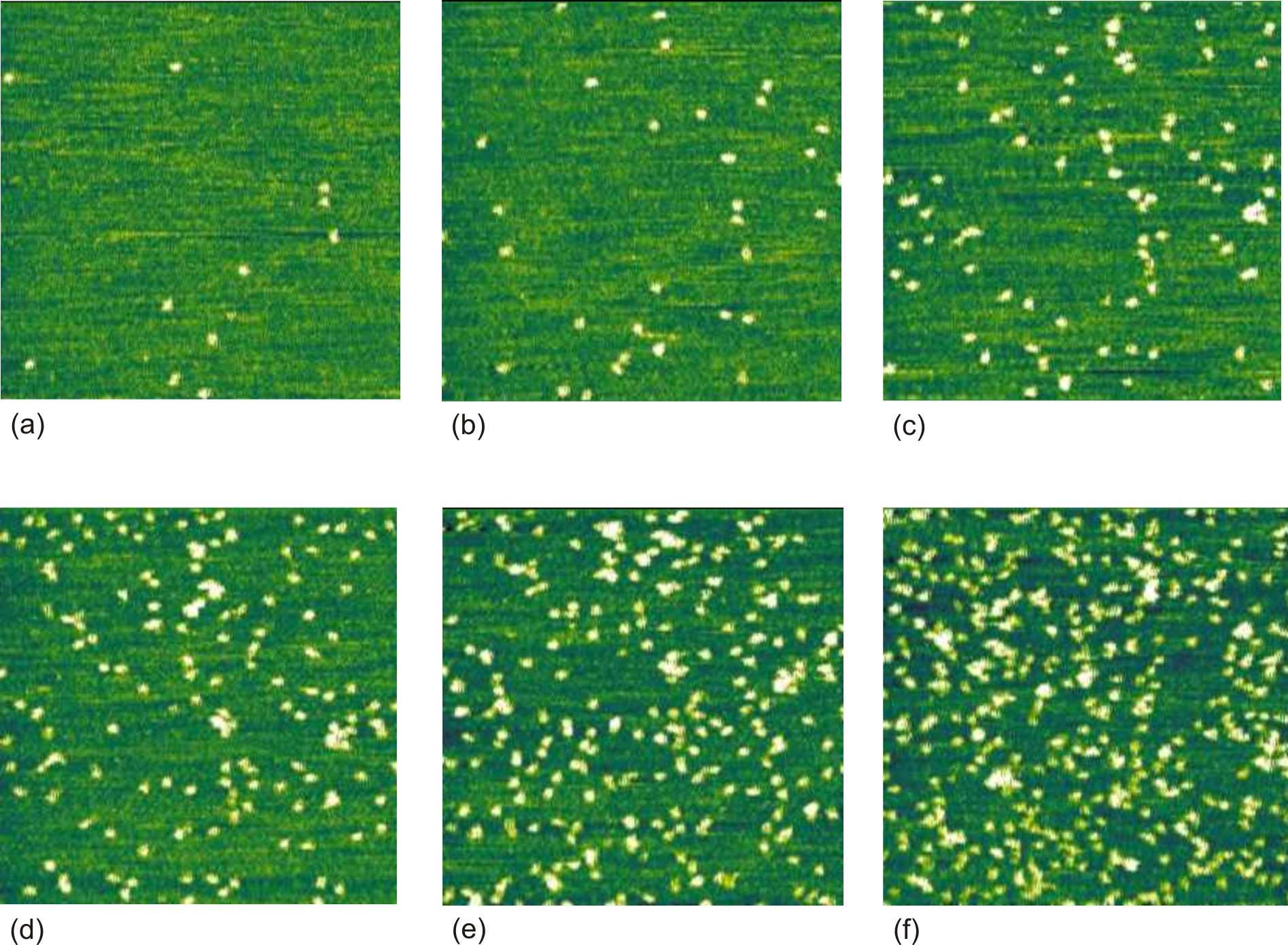}
        \caption{\label{fig:BSAonMica1}Series of consecutive AFM images (1\,$\mu$m\,$\times$\,2$\mu$m) of BSA (20\,$\mu$l of 0.1\,mM protein solution in acetate buffer of pH 4.75 and ionic strength I=8\,mM at room temperature) adsorbed onto mica taken at the same area (slight drift) under stopped flow conditions. The bright objects represent single proteins.}
\end{figure}

By \textit{in situ} scanning, we avoid artifacts arising from the use of linker molecules like glutaraldehyde which typically lead to network-like structures (see Fig. \ref{fig:linkermolekule}(c)) or drying. Both procedures are frequently used in most AFM studies on protein adsorption to enable scanning in air \cite{che2008, ber1998, bel2008, mar2003, agh2006}.

The single objects imaged on the mica can be assigned to single proteins by their dimensions. The bright objects in Fig. \ref{fig:BSAonMica1} have an ellipsoidal shape with a long axis of 22$\pm$7\,nm and a short axis of 15$\pm$5\,nm (without tip deconvolution: nominal tip radius 15\,nm). These dimensions are in good agreement with the dimensions reported in the literature \cite{mal2003} (14 $\times$ 4 $\times$ 4\,nm$^{3}$) indicating that single BSA molecules are displayed on the surface. The height of the proteins ranges between 0.7 and 1.6\,nm depending on the scanning parameters in TappingMode\textsuperscript{\texttrademark} AFM. Therefore it is not a reliable measure for such small and soft objects.

For the measurements on mica, no surface mobility was detected. The interaction between the negativly charged mica and the BSA is believed to hinder the mobility of the proteins. This attractive forces could be screened by solutions of higher ionic strenght. However, reducing the attraction between the proteins and the surface leads to fragile scanning conditions, as mentioned above.

The proteins were not found to preferentially form clusters on the surface but rather form a random distribution \cite{Mecke_et_al}. To objectively characterize the distribution of these proteins on the surface, a two dimensional Minkowski analysis will be used. The Minkowski measures (in two dimensions: covered area, boundary length and Euler characteristics) are quantities that integrate over all n-point correlation functions. This leads to an excellent signal to noise ratio and allows for the structural analysis also for a small number of samples. The Minkowski analysis has successfully been applied to characterize point patterns in biology \cite{mec2005}, astronomy \cite{mec1994} and dewetting processes \cite{her1998}. The knowledge of the statistics of adsorption sites will open an opportunity to determine the interaction potential between the adsorbed proteins and their surface mobility by comparison of experimental results and simulations.

\section{Summary and Conclusions}

The experimental and theoretical characterization of protein biofilms
is a challenging task. In this work we used a number of different
methods in order to characterize the adsorption kinetics of proteins
at different kinds of surfaces. The experimental results  showed that
both the long- and the short-ranged interactions influence strongly the
adsorption kinetics of the proteins.  By means of
ellipsometry, it has been shown that flexible proteins as
e.g. amylase and BSA show an intermediate linear regime upon adsorption
on Si wafers with thin oxide layer.
Complementary computer simulations, which describe the proteins as
single spherical particles with an internal degree of freedom
indicate that this kind of kinetics is the result of conformational
changes induced by density-density fluctuations.
This scenario has been tested experimentally by using different
kinds of proteins and substrates. Yet the adsorption of  more rigid proteins -- like lysozyme -- led to a Langmuir-like adsorption kinetics on all Si wafer types.

A future challenge is to \textit{in situ} study the adsorption site statistics as the surface coverage increases, which enables a direct comparison of experimental and theoretical results. A proof-of-concept is given by \textit{in situ} AFM scans of BSA on mica. Future experiments will explore regimes of different surface mobility of the adsorbing molecules.

To sum up, three points are important for future biofilm studies: i) Van der Waals forces must be taken into account (therefore a detailed characterization of samples is mandatory), ii) conformational changes upon adsorption can be triggered by particle-particle and by particle-surface interactions (where long-ranged forces have to be taken into account, too) and iii) colloidal simulations are currently the only method of choice for describing the kinetics of these systems, which spans over minutes to hours.


%

\end{document}